\documentclass{article}

\usepackage{amsfonts,xr,graphicx,amsmath,amssymb}

\begin{document}
\title{{Liouville theorem for Beltrami flow} }
 \author{{Nikolai Nadirashvili\thanks{Aix-Marseille Universit\'e, CNRS, I2M, 39 rue F. Joliot-Curie, 13453
Marseille  FRANCE, nicolas@cmi.univ-mrs.fr \newline 
The author was partially supported by Alexander von Humboldt Foundation} }}

\date{}
\maketitle

\def\C{\mathbb{C}}
\def\S{\mathbb{S}}
\def\Z{\mathbb{Z}}
\def\R{\mathbb{R}}
\def\N{\mathbb{N}}
\def\H{\mathbb{H}}
\def\tilde{\widetilde}
\def\epsilon{\varepsilon}

\def\n{\hfill\break} \def\al{\alpha} \def\be{\beta} \def\ga{\gamma} \def\Ga{\Gamma}
\def\om{\omega} \def\Om{\Omega} \def\ka{\kappa} \def\lm{\lambda} \def\Lm{\Lambda}
\def\dd{\delta} \def\Dl{\Delta} \def\vph{\varphi} \def\vep{\varepsilon} \def\th{\theta}
\def\Th{\Theta} \def\vth{\vartheta} \def\sg{\sigma} \def\Sg{\Sigma}
\def\bendproof{$\hfill \blacksquare$} \def\wendproof{$\hfill \square$}
\def\holim{\mathop{\rm holim}} \def\span{{\rm span}} \def\mod{{\rm mod}}
\def\rank{{\rm rank}} \def\bsl{{\backslash}}
\def\il{\int\limits} \def\pt{{\partial}} \def\lra{{\longrightarrow}}
\def\pa{\partial } 
\def\ra{\rightarrow }
\def\sm{\setminus }
\def\ss{\subset }
\def\ee{\epsilon }

 {\em Abstract.} We prove that the Beltrami flow of ideal fluid
 in  $\R^3$ of a finite energy is zero.

\bigskip

\section{Introduction}
\bigskip

Let $v(x),\, x\in \R^n,\, n=2,3$ be a velocity of a steady flow of an ideal fluid. Then $v$
is a solution of the system of Euler equation: 

\begin{equation}\label{1} \left\{
\begin{array}{l l}
v \nabla v +\nabla p =0, &\mbox{in $ \R^n $} \\
div\ v=0 &\mbox{in  $ \R^n$} \\
\end{array} \right. \end{equation}

We assume that the vector field $v$ is smooth.

The system of Euler equations  \eqref{1} has equivalent forms. It  can be written as
the Hemholtz equation, see, e.g., [ AK ].

$$
[v,\omega ]=0,
$$
 where $\omega = $curl$ v$ be the vorticity of $v$ and $[ \cdot ]$ be the Lie brackets of
 vector fields. 
 
 In dimension $3$ the Euler equation can be also written  in Bernoulli form:
 
 \begin{equation}\label{3} 
\begin{array}{l l}
v\times \text{curl} v=\nabla b,
 \end{array}
 \end{equation}
 where 
 \begin{equation}\label{4} 
\begin{array}{l l}
b=p+\frac 12 ||v||^2
 \end{array}
 \end{equation}
 be the Bernoulli's function.
 
 A stationary solution $v$ of the system \eqref{1} called the Beltrami flow if $b\equiv $const and
 hence $v$ satisfies the equation
  \begin{equation}\label{2} 
\begin{array}{l l}
v\times \text{curl} v=0
 \end{array}
 \end{equation}

The Beltrami flows is an important class of stationary solutions of the Euler equation. For basic
properties of the Beltrami flows see [AK], some recent results are in  [EP1].

In this paper we are concerned with vanishing at infinity solutions of  \eqref{1}. On the plane the ready example of compactly supported solution of the Euler equation comes from rotationally symmetric flows.
Non-symmetric flow one can obtain pasting together finite or countable collection of rotationally
symmetric flows with disjoint supports.

In dimension $3$ the existence of compactly supported stationary solutions of the Euler equation is not known. However,  there exists a Beltrami flow $v\in C^{\infty }(\R^3)$ such that $|v(x)|<C/|x|$, [EP1].

Notice that nonzero solutions  of   \eqref{1} in $\R^3$  can vanish on an open set,  for instance, the cylinders of  solutions of  \eqref{1} in $\R^2$ with compact support. Explicit
examples of a solution of the Euler equation which vanishes in the interior or exterior part of a given 
hyperboloid constructed in [SV]. In the contrast for the Beltrami flows the unique continuation
property holds, [EP2].

In this paper we show that the Beltrami flow of ideal fluid
 in  $\R^3$ of a finite energy is zero.

\medskip

{\bf Theorem.} {\it  Let $v\in C^1(\R^3)$ be a Beltrami flow. Assume that either $v\in  L_p(\R^3), \,
2\leq p\leq 3$, or
$v(x) = o(1/|x|)$ as $x\to \infty$.
Then $v\equiv 0$.}

\medskip

Notice, that Enciso and Peralta-Salas example of the Beltrami flow, [EP1], shows that the
assumptions of Theorem are sharp.

If we consider the Navier-Stokes equations instead of the Euler equations  then
stronger Liouville type theorems hold. Any bounded in $\R^2$ solution $u$ of the Navier-Stokes equations 
is a constant, see [KNSS], and any solution of the Navier-Stokes equations in $\R^3$ with a 
sufficiently small $L_3$-norm is zero, see [G].

To prove Theorem 1.2 we rewrite equations  \eqref{1} as linear equations for a suitable tensor form.

\section{Tensor equations from the Euler equation}

First we introduce some tensor notations and then derive from \eqref{1} equations for corresponding tensor
 fields.

 Denote by $T^m$ the space of covariant tensors on $\R^n$ of the rang $m$; $S^m\ss T^m$
 be the symmetric subspace of $T^m$. The map $\sigma :T^m\to S^m$
 $$\sigma f(x_1,\dots,x_m)= \frac1{m!}\sum   f(x_{i_1},\dots,x_{i_m})       $$
 where the summation is taken over all permutations of the indices $1,É,m$,
 called the symmetrization of tensor $f$. For smooth tensor fields $C^{\infty }(T^m,\R^n)$ is
 defined covariant differentiation $\nabla : C^{\infty }(T^m,\R^n)\to C^{\infty }(T^{m+1},\R^n)$,
 $$\nabla f =  f_{i_1,\dots,i_m; j}$$
 The operator $d$ of inner differentiation is the symmetrization of $\nabla$,
 $d=\sigma\nabla : C^{\infty }(S^m,\R^n)\to C^{\infty }(S^{m+1},\R^n)$. The divergence operator
 $\delta $,  $\delta: C^{\infty }(S^m,\R^n)\to C^{\infty }(S^{m-1},\R^n)$,
 $$(\delta f)_{i_1,\dots,i_{m-1}}=\sum f_{i_1,\dots,i_m; i_m}$$
 is an operator formally adjoint to $-d$.
 
 Let $v\in C^{\infty}(R^3)$ be a solution of \eqref{1}. We define the tensor $F\in   C^{\infty}(S^2,R^3) $ of the flow $v$ as
 $$F=p(dx)^2+\tilde v^2,$$
 where $\tilde v$ is a convector dual to the vector $v$: $\tilde v(\cdot )= (v, \cdot)$ and 
 $$\tilde v^2=\sum v^iv^jdx_idx_j.$$
 
 As a consequence of the system \eqref{1} one has the equations
 $$p_i+\sum_j (v^iv^j)_{j}=0$$

Directly from the last equations we get the following linear equation for $F$:
\begin{equation}\label{5}
 \delta F=0
  \end{equation}

\section{Proof of the theorem}

For a Beltrami flow $v$ it follows from \eqref{2}, \eqref{3}, \eqref{4} that $p=-|v|^2/2+$const.
Subtracting from $p$ a constant we may assume that
\begin{equation}\label{6}
p=-|v|^2/2
 \end{equation} 

Let $F$ be the flow's tensor of $v$. Then from \eqref{6} it follows
$$F=-{|v|^2\over 2}(dx)^2+{\tilde v}^2,$$
where $(dx)^2=(dx_1)^2+(dx_2)^2+(dx_3)^2$.

Let $A$ ($B$) be the the spherical average of $F$ ($\tilde v$), i.e.,

$$A=\int_{s\in O_3}F_sd\chi,$$
$$B=\int_{s\in O_3}{\tilde v}^2_sd\chi$$
where $F_s$ (${\tilde v}_s$) are the rotations of  $F$ (${\tilde v}$) on $s\in O_3$ and $d\chi $
be the Haar measure on the group $O_3$. Then
$$tr\, A(x)= -\frac 12 tr\, B(x)$$
and hence
$$A(x)=B(x)-\frac12(tr\, B(x))(dx)^2$$

 Let $r\in \R,\, \theta \in S^2$ be the polar coordinates in $\R^3$, $r^2drd^2\theta$ is a standard element  of volume in $\R^3$:   $r^2drd^2\theta=(dx)^3$, where $d^2\theta$ is the area form
 on the unit sphere.  Let
$$B=\alpha (x)(dr)^2+ \beta (x)r^2(d\theta)^2,$$
where $(d\theta)^2$ is the metric tensor of the unit sphere and$\alpha (x)=\alpha (|x|),\, \beta (x)=\beta (|x|)$. Since $\tilde v^2$ and hence $B$ are nonnegative tensors then $\alpha,\beta\geq 0$. Therefore
$$A(x)=B(x)-\frac12(\alpha +2\beta)(dx)^2= (\frac12\alpha -\beta)(dr)^2 -\frac12\alpha(d\theta)^2     $$

Since \eqref{5} is a linear equation it holds after the averaging of $F$,
\begin{equation}\label{7}
\delta A=0
 \end{equation} 
 
 Denote by $G_r$ the half ball $\{ \{ |x|<r\}\cap \{ x_1<0\}\}$. Set  $H_r=\{ \{ |x|<r\}\cap \{ x_1=0\}\}$, $n=(1,0,0)$. Integrating equality \eqref{7}
 against the vector $n$ we get
 $$- \int_{H_r}( An,n)ds= \int_{\partial G_r\sm H_r}(An,x/r)ds$$
 Since  $(An,n)_{|\{x_1=0\}}=-\alpha /2 \leq 0$ we have
  \begin{equation}\label{9}
\int_0^rt\alpha(t)dt =-\frac12r^2(\alpha ( r)-2\beta ( r))
  \end{equation} 
 Hence
 \begin{equation}\label{8}
- \int_{H_r}( An,n)ds\leq \int_{\partial G_r\sm H_r}|A|ds
  \end{equation} 
By our assumption either $v\in L_p(\R^3), \, 2\leq p\leq 3$, and hence
  \begin{equation}\label{10}
\int_0^{\infty}\int_{\partial G_r\sm H_r}|A|^{p/2}dsdr< \infty
  \end{equation} 
 or $v(x) = o(1/|x|)$ as $x\to \infty$, therefore  $|A|= o(1/|x|^2)$ and
   \begin{equation}\label{11}
 \int_{\partial G_r\sm H_r}|A|ds = o(1/ r ),
   \end{equation} 
   In the first case by H\" older's inequality
   $$\int_{\partial G_r\sm H_r}|A|ds\leq \left( \int_{\partial G_r\sm H_r}|A|^{p/2}ds\right)^{2/p}
\left( \int_{\partial G_r\sm H_r}ds\right)^{(p-2)/p}  $$
or
   \begin{equation}\label{12}
  \left(\int_{\partial G_r\sm H_r}|A|ds\right)^{p/2}\leq 2\pi r^{p-2}\int_{\partial G_r\sm H_r}|A|^{p/2}ds   \end{equation} 
  From inequality \eqref{10} follows the existence of the sequence $r_n \to \infty$ such that
 $$r_n \int_{\partial G_{r_n}\sm H_{r_n}}|A|^{p/2}ds \to 0   $$
 as $n\to \infty$.
 Thus from the inequality \eqref{12} follows that
  $$\int_{\partial G_{r_n}\sm H_{r_n}}|A|ds \to 0   $$
 as $n\to \infty$.

 Since $\alpha $ is nonnegative  taking $n\to \infty$ we get from the inequality \eqref{8} 
 $$\alpha \equiv 0$$
 in  case of inequality \eqref{10}. In case  \eqref{11} the last identity immediately follows from \eqref{8} and \eqref{11}. 
 Then from the equality  \eqref{8} we conclude
 $$\beta \equiv 0$$

Thus   $A=0$ and hence
 $v(0)=0$. Since the last equality holds for the any choice of origin in $\R^3$ it
 follows that $v\equiv 0$. The theorem proved.

\bigskip

\bigskip

\bigskip
 \centerline{REFERENCES}

 \medskip
\noindent [AK] V.I. Arnold, B.A. Khesin {\it Topological Methods in Hydrodynamics },
Sringer, 1998.

 \medskip
\noindent [G] G.P. Galdi {\it An Introduction to the Mathematical Theory of the Navier-Stokes 
Equations Volume 2}, Springer  1998

  \medskip
\noindent [EP1] A. Enciso, D. Peralta-Salas  {\it   Existence of knotted vortex tubes in steady Euler flows,}
 arXiv:1210.6271v1
 
   \medskip
\noindent [EP2] A. Enciso, D. Peralta-Salas  {\it Beltrami fields with a non constant proportionalty
factor are rare }, preprint

 \medskip
\noindent [KNSS] G. Koch, N. Nadirashvili, G. Seregin, V. Sverak {\it Liouville theorems for the Navier-Stokes equations and applications. }Acta Math. 203 (2009), no. 1, 83-105.

 \medskip
\noindent [SV]  B. Shapiro, A.Vainstein {\it Multididimensional analogues of the Newton and
Ivory theorems}, Funct. Anal. Appl. 19 (1985), 17-20

\end{document}